\RequirePackage{lineno}
\documentclass[twocolumn,showpacs,preprintnumbers,amsmath,amssymb,floatfix,superscriptaddress]{revtex4}
\usepackage{graphicx}% Include figure files
\usepackage{dcolumn}% Align table columns on decimal point
\usepackage{bm}% bold math
\usepackage{mathrsfs}
\usepackage{color}
\usepackage{ulem}

\def\mcR{\mathcal{R}}

\def\mcF{\mathcal{F}}
\def\mcJ{\mathcal{J}}

\begin{document}

\title{
\begin{flushright}
{\small \sl version 2.5,  \today \\
 }
\end{flushright}
The Study of Noncollectivity by \\
the Forward-Backward Multiplicity Correlation Function \\
}% Force line breaks with \\

\date{\today}% It is always \today, today,

\affiliation{Department of Physics, Huazhong University of Science and Technology, Wuhan 430074, China}
\affiliation{Institute of Particle Physics, CCNU (HZNU), Wuhan 430079, China}
\affiliation{Brookhaven National Laboratory, Upton, New York 11973, USA}
\author{Na Li}\email{nli@mail.hust.edu.cn}\affiliation{Department of Physics, Huazhong University of Science and Technology, Wuhan 430074, China}
\author{Shusu Shi}\email{sss@iopp.ccnu.edu.cn}\affiliation{Institute of Particle Physics, CCNU (HZNU), Wuhan 430079, China}
\author{Aihong Tang}\affiliation{Brookhaven National Laboratory, Upton, New York 11973, USA}
\author{Yuanfang Wu}\affiliation{Institute of Particle Physics, CCNU (HZNU), Wuhan 430079, China}

\begin{abstract}
We propose a forward-backward multiplicity correlation function $C^N_{FB}$, which is experimentally accessible, to measure the noncollectivity contribution.
It is found that the function is sensitive to both of the jet contribution and the small number statistics. We point out that the effect of the latter one is also involved in the previous studies of the forward-backward elliptic correlation function but was confused as the contribution from jets. We study the $C^N_{FB}$ in Au+Au collision at $\sqrt{s_{NN}}=200$ GeV with a multiphase transport model (AMPT). The result shows that the estimated jet contribution is much smaller than previous study due to the finite number statistics which has not been noticed before. The connection between this study and the forward-backward elliptic correlation function is also discussed.

\end{abstract}



\maketitle

\setpagewiselinenumbers
\modulolinenumbers[1]
%\linenumbers

\section{Introduction}
Quantum Chromodynamics predicts a new form of matter, a deconfined state of quarks and gluons called quark-gluon plasma (QGP). It can be created by heavy ion collisions at relativistic energies experimentally~\cite{QGP}. In a collision, particles are produced by two main processes: soft hadrons are generated from the hot dense medium created by energy sedimentation, while jets are produced from initial hard scatterings.  Being relevant to the two processes respectively, the discovery of the large elliptic anisotropy ($v_2$)~\cite{QGP, v2_star, PHOBOS2002} and the jet quenching~\cite{quenching} have been taken as important evidences of QGP at the Relativistic Heavy Ion Collider (RHIC)~\cite{RHIC}.

Particles produced by different processes carry different information of the system. For example, the soft hadron azimuthal anisotropy at low transverse momentum, called elliptic flow, is formed due to the hydrodynamic pressure buildup in the initial almond overlap region of the colliding nuclei~\cite{NQScaling, v2_star}. At large transverse momentum, energetic partons are predicted to lose energy by induced gluon radiation~\cite{quenching}. This energy loss depends strongly on the traversed path length of the propagating parton, thus also leads to azimuthal anisotropy $v_2$~\cite{Snellings:1999gq, Wang:2000fq}. The $v_2$ measured from final state particles is the combination of these two effects. However, the elliptic flow $v_2$ and azimuthal anisotropy $v_2$ of jets individually might not be the same. Trying to separate the two, previous work~\cite{jingfeng} in this regard has focused on the covariance of $v_2$ from forward and backward regions, which has been found later that the method is complicated by $v_2$ fluctuations~\cite{han}, and the jet contribution still can not be directly accessed from final state particles. In this paper, we approach the problem differently by studying the multiplicity correlation.

\section{The forward-backward multiplicity correlation function}
The procedure is similar to the previous study~\cite{jingfeng}. To insure that jets only contribute to one region while collective particles contribute to both in a single event, two regions with the same size in phase space are chosen. Since jets are produced locally, following Ref.~\cite{jingfeng}, we make forward-backward rapidity bins with a gap between them. The proposed new observable $C_{FB}^N$ is defined as the standard correlation function:
\begin{equation}
C_{FB}^{N}\equiv \frac{\langle N_F N_B \rangle}{\langle N_F\rangle\langle N_B\rangle}-1,
\label{CFB}
\end{equation}
where $N_{F(B)}$ refers to the particle yields in the forward (backward) region, and $\langle\cdots\rangle$ means taking the average over events.

\begin{figure}
\label{fig:fig1}
\centering
\includegraphics[width=3.2 in]{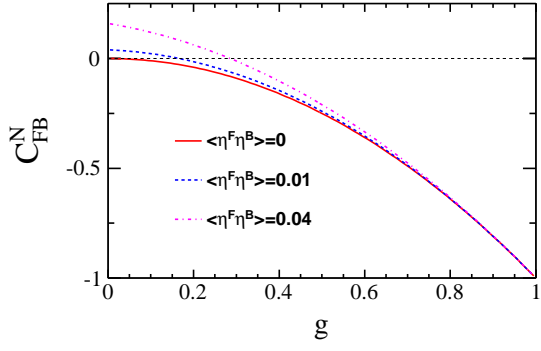}
\caption{\label{fig:fig1} (Color online) The dependence of forward-backward multiplicity correlation function $C_{FB}^N$ on jet fraction $g$.}
\end{figure}

To count the contribution of jets to the total multiplicity, we define a $g$-parameter:
\begin{equation}
g = \frac{\langle N^{\mathcal{J}} \rangle}{\langle N^{\mathcal{F}}+N^{\mathcal{J}}\rangle}.
\label{gfun}
\end{equation}
Here $N^{\mathcal{F(J)}}$ refers to the multiplicity of flow (jets). Since the experimental efficiency does not bias flow or jet particles, forward or backward regions, it is not expected to affect the measurement. Similar to~\cite{jingfeng}, in each event,  the yields in forward (backward) region can be written as
\begin{align}
N_F &= (\frac{1}{2}+\eta^F)\langle N^{\mathcal{F}}\rangle + \xi N^{\mathcal{J}}, \notag \\
N_B &= (\frac{1}{2}+\eta^B)\langle N^{\mathcal{F}}\rangle + (1-\xi)N^{\mathcal{J}},
\label{nfun}
\end{align}
with
\begin{align}
N=N_F+N_B=N^{\mathcal{F}}+N^{\mathcal{J}}.
\end{align}
In the above, $N$ is the total multiplicity in an event, and $\eta^{F(B)}$ is the event-by-event multiplicity fluctuation of flow particles in each region. The jet multiplicity may also fluctuates event-by-event, thus no average is taken for $N^\mathcal{J}$.  After taking the average over all events, we will get $\langle\eta^{F(B)}\rangle=0$. For the final state particles, the correlation between flow particles and jets is believed to be small thus ignored, and this is justified by that they are produced at different moments during the collision, i.e., $\langle N^{\mathcal{F}}N^{\mathcal{J}}\rangle = \langle N^{\mathcal{F}}\rangle\langle N^{\mathcal{J}}\rangle$. $\xi$ describes the jet contribution to the forward and backward region. Since jets only fall in one side in an event, the assumed value is 1 or 0. The location is randomly decided, so the probability is the same for each side, i.e.,$\langle\xi\rangle=\langle1-\xi\rangle=1/2$ while $\xi(1-\xi)$=0.

Combine Eq.~(\ref{gfun}) and (\ref{nfun}) and insert into  Eq.~(\ref{CFB}), we find the relation for $g$-parameter and $C_{FB}^N$:
\begin{align}
%g = \sqrt{-C^N_{NA}}.
C_{FB}^N &= - g^2 + 4\langle\eta^F\eta^B\rangle(1-g)^2. %\notag \\
%   &+\frac{2(\langle M^\mcF M^\mcJ\rangle-\langle M^\mcF\rangle\langle M^\mcJ\rangle)}{\langle M^\mcF+M^\mcJ\rangle^2} .
%    &+\frac{2\langle(\eta_1+\eta_2)M^\mcJ\rangle\langle M^\mcF\rangle}{\langle M^\mcF+M^\mcJ\rangle^2}.
\label{C2}
\end{align}
$g$ is the contribution of jets and $\langle\eta^F\eta^B\rangle$ is a measure of correlated forward-backward fluctuations. Both of them affects the behavior of $C_{FB}^N$.

Figure~\ref{fig:fig1} shows the relation between $C_{FB}^N$ and the $g$-parameter. In the ideal case, if flow particles from the two regions fluctuate independently, i.e., $\langle\eta^F\eta^B\rangle=0$, as assumed in Ref~\cite{jingfeng}, Eq.~(\ref{C2}) can be simply written as
\begin{equation}
C_{FB}^N=-g^2.
\label{C3}
\end{equation}
Then, $g$ can be extracted from the value of $C_{FB}^N$.
If the system is hydro dominated, $g\rightarrow0$ and $C_{FB}^N\rightarrow0$. While if the system is jets dominated, $g\rightarrow1$ and $C_{FB}^N\rightarrow -1$. On the other hand in a realistic case, if flow particles from the two regions do not fluctuate independently ($\langle\eta^F\eta^B\rangle\neq0$), when $g$ is small, $C_{FB}^N$ will be larger than 0, and when $g$ is large, the effect of $\langle\eta^F\eta^B\rangle$ can be ignored.

\begin{figure}
\centering
\includegraphics[width=3.2 in]{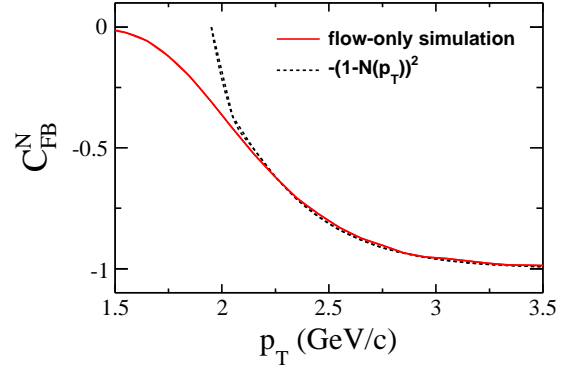}
\caption{\label{fig:fig2}(Color online) $C_{FB}^N$ as a function of $p_T$ for flow-only simulation. The black dash curve is from Eq.~(\ref{C4}).}
\end{figure}

The Eq.~(\ref{C2}) works in the low $p_T$ with relatively large number statistic. However, if there is only one flow particle produced a particular $p_T$ bin in a event, it can only fall into one side and behaves like a jet. In that case, the multiplicity correlation function will automatically decrease when the particle yield is much smaller than 1. This effect has not been discussed in previous studies~\cite{jingfeng, han}, and can be easily confused with the contribution from jet. To explore this effect, a simulation was performed as followed. The flow particles were generated with a exponentially decreases distribution, and randomly fell into forward/backward region. Adopting the consistent terminology as in ~\cite{jingfeng}, we call them "flow-only" particles throughout this paper. The red curve in Fig.~\ref{fig:fig2} shows the result of flow-only simulation. It can be seen that although there is no jet, $C^N_{FB}$ is still approaching to $-1$ at large $p_T$. This phenomena is solely due to, for a given $p_T$ bin, the exclusion of events for which both $N_F$ and $N_B$ for that $p_T$ bin are zero, from the $C_{FB}$ calculation. This happens even when the two samples are completely independent of each other.

With the simplest assumption that when particle yield $N(p_T)$ is smaller than 1, maximum one particle can be found in either forward or backward region, the probability of finding an empty $p_T$ bin is $(1-N(p_T)/2)^2$. If one calculates $\frac{\langle N_F N_B \rangle}{\langle N_F\rangle\langle N_B\rangle}$ without the exclusion of events with empty bins as mentioned above, for two independent samples it should be at unity. Considering the exclusion, one needs to scale each averaging term in $\frac{\langle N_F N_B \rangle}{\langle N_F\rangle\langle N_B\rangle}$ by a factor of $\frac{1}{1 - (1-N(p_T)/2)^2}$, then $C^N_{FB}(p_T)$ can be written as:
\begin{equation}
C_{FB}^N(p_T)=-(1-N(p_T)/2)^2.
\label{C4}
\end{equation}
Thus even for independent samples, with the exclusion of empty bins, $C^N_{FB}(p_T)$ is decreasing with increasing $N(p_T)$. The black dash line in Fig.~\ref{fig:fig2} is a plot of Eq.~(\ref{C4}). We can see that when $p_T$ is larger than 2.2 GeV/$c$, the equation can well describe the simulation result. While when $p_T$ is smaller than 2.2 GeV/$c$, the simulation result is lower than the black dash curve. This is because when particle yield is not so small, there might be more than one particle generated in the forward/backward region for the flow-only simulation. Therefore, from the Eq.~(\ref{C3}) and Fig.~\ref{fig:fig2} we can conclude that both of the jets and the scarcity of particles can cause the decreasing trend of $C^N_{FB}(p_T)$. We may also expect that it will decline even faster when jets are involved in the particle-rare case. %Thus, with the same yield, the probability of finding empty bins in the flow-only simulation is larger, and $C^N_{FB}(p_T)$ is smaller.

\section{AMPT model}
The forward-backward multiplicity correlation function $C^N_{FB}$ is studied in a multiphase
transport model (AMPT)~\cite{AMPT}. There are four main components in this transport model:
the initial conditions, the parton-parton interactions, the conversion from the partonic to the
hadronic matter and the late hadronic interactions. The initial conditions
are based on the HIJING model~\cite{HIJING} in which the eikonized parton model is employed.
It includes the spatial and momentum distributions of minijet partons from hard processes and
strings from soft processes.
The parton-parton interactions and the time evolution of partons is then treated
by the Zhang's Parton Cascade (ZPC)~\cite{ZPC} model.
The hadronization process is described by a combined coalescence and string fragmentation model.
A relativistic transport (ART) model~\cite{ART} which includes baryon-baryon, baryon-meson and meson-meson elastic and inelastic scattering
is employed to describe the late hadronic process.
In our study, we analyzed the events from string melting AMPT model for Au + Au collisions at $\sqrt{s_{NN}}=200$
GeV with parton cross sections equal to 10 mb. Pseudorapidity regions $[-1,-0.5]$ and  $[0.5,1]$
are chosen as forward and backward region respectively.

\begin{figure}
\centering
\includegraphics[width=3.2 in]{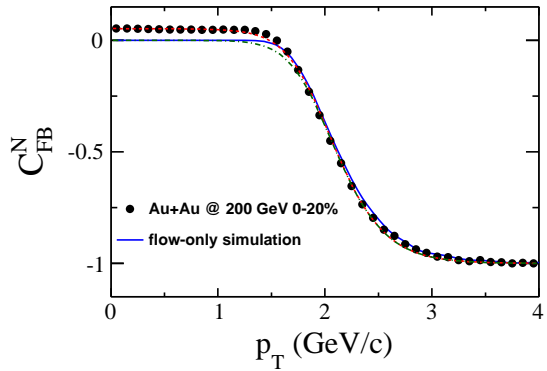}
\caption{\label{fig:fig3}(Color online) $C_{FB}^N$ as a function of $p_T$ for Au+Au collision at $\sqrt{s_{NN}}$=200 GeV in AMPT string melting model. The red dash curve is the fitting of data with the form of Eq.~(\ref{fit1}), and the green dash-dot curve is a plot of Eq.~(\ref{fit2}). The blue curve is the flow-only simulation. }
\end{figure}

\begin{figure}
\centering
\includegraphics[width=3.6 in]{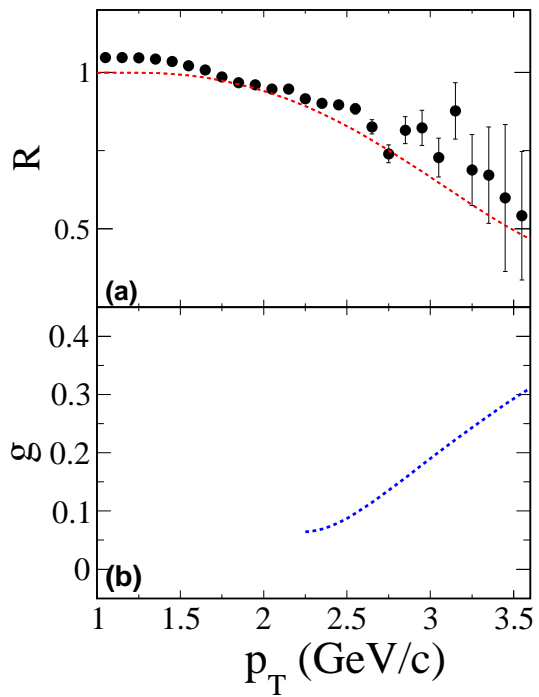}
\caption{\label{fig:fig4}(Color online)(a) ($1+C^N_{FB}$) (solid points) and ($1+C'^N_{FB}$) (red dash curve) from AMPT are divided by ($1+C^N_{FB}$) from flow-only simulation. (c) $g$-parameter as a function of $p_T$.}
\end{figure}

The $p_T$ dependence of multiplicity correlation function is shown in Fig.~\ref{fig:fig3}. We can see that $C_{FB}^N$ almost stays as a constant when $p_T\leq1.5$ GeV/$c$. Since particles from high energy jets give only a small contribution at those $p_T$, the value of $C_{FB}^N$ reflects the flow multiplicity fluctuation, $4\langle\eta^F\eta^B\rangle$. Although the particle yield decreases exponentially, $4\langle\eta^F\eta^B\rangle$ shows little $p_T$ dependence for this region. To obtain the jet contribution, $C_{FB}^N$ is fitted by the function:
\begin{equation}
%C_{FB}^M = - \{1+\tanh[(p_T-P_C)/P_W]\}^2/4 + a_0
C^N_{FB}=-q(p_T)^2+4\langle\eta^F\eta^B\rangle(1-q(p_T))^2,
\label{fit1}
\end{equation}
and $q(p_T)$ is parameterized using a tanh function to describe threshold behavior between the two regimes, as suggested in ~\cite{jingfeng}:
\begin{equation}
%C_{FB}^M = - \{1+\tanh[(p_T-P_C)/P_W]\}^2/4 + a_0
q(p_T)=\{1+\tanh[(p_T-P_C)/P_W]\}/2.
\label{fit}
\end{equation}
Here $q(p_T)$ is not simply $g(p_T)$ anymore. Instead, it is a convolution of both $g(p_T)$ (dominated at low $p_T$) and the automatic decrease of $C_{FB}$ (dominated at large $p_T$). After fitting, we could obtain the first part of Eq.~(\ref{fit1}):
\begin{equation}
%C_{FB}^M = - \{1+\tanh[(p_T-P_C)/P_W]\}^2/4 + a_0
C'^N_{FB}=-q(p_T)^2,
\label{fit2}
\end{equation}
where the effect of flow multiplicity correlation between the forward and backward regions has been canceled.

In Fig.~\ref{fig:fig3}, we present the fitting of AMPT data by the red dash curve. We can see that this form can well describe the trend of the data, and we obtain that $\langle\eta^F\eta^B\rangle\approx0.013$. For comparison, we fit the particle distribution with Levy function, which can well describe the data, and generates flow particles with the same distribution. As discussed before, the flow particles will randomly fall into each side in this simulation. The result of this flow-only simulation is shown as the blue curve in Fig.~\ref{fig:fig3}. From the plot, we can see that the simulation stays zero at low $p_T$ as expected since there is neither jet nor correlation. While at intermediate $p_T$ range, both of the AMPT data and the flow-only simulation decreases as $p_T$ increasing.  We found that the AMPT result including jet contribution is only slightly lower than the flow-only simulation. Therefore, we argue that the decreasing trend of the forward-backward correlation function as shown in~\cite{han} may not only due to the jet contribution, but mostly due to the artificial decrease for the particle-rare case.

To see the details clearly, the results from AMPT model are divided by that from the flow-only simulation. Here we define
\begin{equation}
\mcR=\frac{\langle N_F N_B \rangle_{AMPT}}{\langle N_F N_B \rangle_{Simulation}}=\frac{(C^N_{FB}+1)_{AMPT}}{(C_{FB}^N+1)_{Simulation}}
\label{fit}
\end{equation}
%Because the particle distribution in the simulation is made as the same as that from AMPT, the term $\langle N_F\rangle\langle N_B\rangle$ cancels.
The deviation from unity of this ratio reflects the dilution of $\langle N_F N_B \rangle_{AMPT}$ due to pairs containing jets. From the plot, we can see that the ratio begins to deviate from 1 around 2 GeV/$c$, which indicates that the jet contribution begins to join in this region. The difference between the points and the curve comes from flow multiplicity correlation.

For qualitatively study, we assume that the square root of the ratio is proportional to the particle yield which is expected to work for $p_T>2.2$ GeV/$c$ as shown in Fig.~\ref{fig:fig2}, and finally $g$ can be extracted from $(1-\sqrt{\mcR})$. Fig.~\ref{fig:fig4}(b) shows that the $g$-parameter is about 10\% at 2 GeV/$c$ and reaches 30\% at 3 GeV/$c$. Our method works best for central collisions in which away-side production of a jet is suppressed thus those particles can be considered as being ``melted'' into flow particles. While for the non-central collision, the $g$ quantity will be reduced due to the not quenched away-side jets. In that case, the correlation function measures the $extra$ jets in one side.

\section{The connection to previous observable $C_{FB}^{V_2}$}
The forward-backward elliptic anisotropy correlation has been discussed in Ref~\cite{jingfeng,han}. This correlation function also subjects to the small number statistic discussed before. Here we redefine the elliptic correlation function:
\begin{equation}
C_{FB}^{V_2}\equiv \frac{\langle V_2^F V_2^B\rangle}{\langle V_2^F\rangle\langle V_2^B\rangle}-1,
\label{CFBv2}
\end{equation}
where $V_2^{F(B)}$ refers to the sum of the $v_2$ of particles in forward (backward) region in each event. The $\langle v_2\rangle$ we measured in experiments is constituted of two parts: flow $v^{\mcF}_2$ and jet $v_2^\mcJ$:
\begin{equation}
\langle v_2\rangle = (1-g)\langle v_2^\mcF\rangle + g\langle v_2^\mcJ\rangle.
\label{v2con}
\end{equation}
$v_2^\mcF$ describe the collective behavior of hydro, and $v_2^\mcJ$ is related to the parton energy loss in different directions. The relation between $C_{FB}^{V_2}$ and pure jet (flow) $v_2$ can then be written as:
\begin{equation}
C_{FB}^{V_2}=-\frac{g^2(v_2^\mcJ)^2}{\langle v_2\rangle^2}+
\frac{4\langle\eta^F\eta^B\rangle(1-g)^2\langle v_2^\mcF\rangle}{\langle v_2\rangle^2}.
\label{v2}
\end{equation}
The differences between this and Ref~\cite{han} is coming from the $-1$ in the definition. One should pay attention that $\langle\eta^F\eta^B\rangle$ here is not the same as in the multiplicity correlation. For multiplicity correlation, $\eta^{F(B)}$ refers to the total multiplicity fluctuation in each side. While for $v_2$ correlation, $\eta^{F(B)}$ have azimuthal dependence, i.e., the in-plane and out-of-plane multiplicity fluctuation may cause the statistic fluctuation of $v_2$. To be clearer, the Eq.~(\ref{v2}) can be written as
\begin{equation}
C_{FB}^{V_2}=-\frac{g^2(v_2^\mcJ)^2}{\langle v_2\rangle^2}+
\frac{4(1-g)^2\sigma_{v_2^\mcF}^2}{\langle v_2\rangle^2}.
\label{v2new}
\end{equation}

Therefore, for the hydro dominate case, i.e., $g\rightarrow0$, $C_{FB}^{V_2}=4\frac{\sigma_{v_2}^2}{\langle v_2 \rangle}$. The $C_{FB}^{V_2}$ describes the $v_2$ fluctuation at low $p_T$, and the large value of the $C_{FB}^{V_2}$ observed in Ref~\cite{han} could be understood according to the magnitude of the flow fluctuation studied in Ref~\cite{Paul}. For intermediate $p_T$, however, the decreasing trend of $C_{FB}^{V_2}$ observed ~\cite{han} are mostly due to the rare flow particles as discussed before. Therefore, it is interesting to check the behavior of $C_{FB}^{V_2}$ in the LHC energy where particle yield is much higher.

\section{Conclusion}
In summary, we have proposed a forward-backward multiplicity correlation function $C^N_{FB}$  to study the jet contribution. We find that the $C^N_{FB}$ is sensitive to the jet contribution, and its value will be enlarged by the forward-backward multiplicity correlation. In the intermediate $p_T$, We find that the $C^N_{FB}$ is sensitive to small number statistic and will automatically decrease even there are no jets involved. This effect should also be present in previous study of $C^{V_2}_{FB}$, and might be misunderstood as the contribution of jets.

We study the $C^N_{FB}$ in Au+Au collisions at $\sqrt{s_{NN}}=200$ GeV with AMPT string melting model. When $p_T$ is lower than 1.4 GeV/$c$, $C^N_{FB}$ is independent of $p_T$, and $\langle\eta^F\eta^B\rangle$ is about 0.01. At intermediate $p_T$ range, $C^N_{FB}$ decreases with $p_T$, and the result from AMPT model is lower than the flow-only simulation of the same yield. It indicates that both of the the jet contribution and the procedure of excluding events with empty bins will cause the decrease of $C^N_{FB}$. The estimated jet fraction is much smaller than
previous study. 

Finally, we discussed the connection of our study to the forward-backward elliptic anisotropy correlation function. The large value of $C^{V_2}_{FB}$ is found due to $v_2$ fluctuation, and the decrease of  $C^{V_2}_{FB}$ observed previously is caused by both of the jet and the tail of flow particle distribution, with the latter dominates.

%\vskip 0.5cm
The authors thank L. X. Han, J. Liao and G. Wang for the discussion. The work was supported in part by Huazhong University of Science and Technology Foundation under grant no. 2011QN195, the National Natural Science Foundation of China under grant no. 11147196, 11105060 and 10835005, and by the Office of Nuclear Physics, US Department of Energy under Grants DE-AC02-98CH10886 and DE-FG02-89ER40531.

\end{document}